# STUDY ON THE INJECTION BEAM COMMISSIONING SOFTWARE FOR CSNS/RCS*

M.Y. Huang[#], S. Wang, W.B. Liu, J. Qiu, L.S. Huang

[1] China Spallation Neutron Source, Institute of High Energy Physics,
Chinese Academy of Sciences, Dongguan, China

[2] Dongguan Institute of Neutron Science, Dongguan, China

*Abstract*

The China Spallation Neutron Source (CSNS) accelerator uses $H^-$ stripping and phase space painting method of filling large ring acceptance with the linac beam of small emittance. The beam commissioning software system is the key part of CSNS accelerator. The injection beam commissioning software for CSNS contains three parts currently: painting curve control, injection beam control and injection orbit correction. The injection beam control contains two subsections: single bunch beam calculation and LRBT beam control at the foil. The injection orbit correction also contains two subsections: injection orbit correction by the calculation and injection trim power control.

## INTRODUCTION

CSNS is a high power proton accelerator-based facility [1]. The accelerator consists of an 80MeV $H^-$ linac and a 1.6GeV Rapid Cycling Synchrotron (RCS) which accumulates an 80MeV injection beam, accelerates the beam to the designed energy of 1.6GeV and extracts the high energy beam to the target. Its beam power is 100kW and capable of upgrading to 500kW. The design goal of CSNS is to obtain the high intensity, high energy proton beam with a repetition rate of 25Hz for various scientific fields [2].

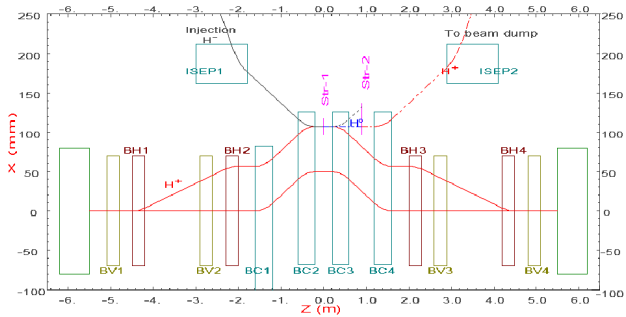

Figure 1: Layout of the RCS injection system.

For CSNS/RCS, a combination of the $H^-$ stripping and the phase space painting method is used to accumulate a high intensity beam. Figure 1 shows the layout of the RCS injection system [3]. For the injection system, three kinds of orbit-bumps are prepared: a horizontal bump (BH1-BH4) for painting in x-x′ plane; a vertical bump (BV1-BV4) for painting in y-y′ plane; a horizontal bump (BC1-BC4) in the middle for an additional closed-orbit shift of 60mm [4].

The beam commissioning software which was programmed with the Java language plays a very important role in CSNS project, and it bases on the XAL application development environment which was developed initially by SNS laboratory [5]. The main application of this beam commissioning software system contains: device control, monitoring, online modeling and data analysis functions [6].

## PAINTING CURVE CONTROL

In order to control the strong space charge effects which are the main causes of the beam losses in CSNS/RCS, the phase space painting method is used for injecting the beam of small emittance from the linac into the large ring acceptance. In general, there are two painting methods: correlated painting and anti-correlated painting. From our simulation results by using the code ORBIT [7], it can be found that the anti-correlated painting method may be more suitable for CSNS/RCS.

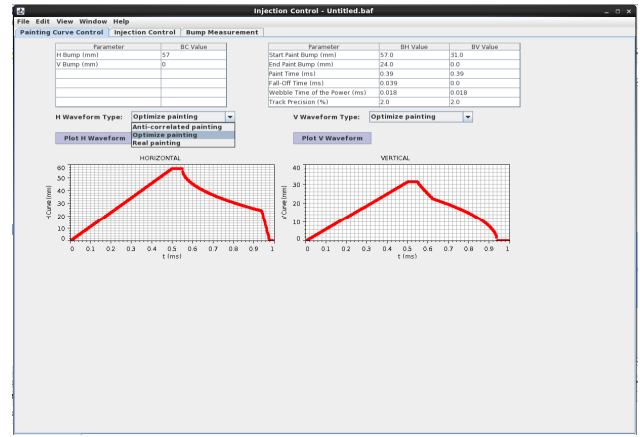

Figure 2: Injection painting curves control.

For the injection beam commissioning software of CSNS/RCS, there are three kinds of painting curves: ideal anti-correlated painting curve, optimize painting curve, and real painting curve. The optimized painting curve is obtained by the simulation and optimization of injection process. The real painting curve will be given by the accelerate operation and test. Figure 2 shows the control interface of the injection painting curves. It can be found that the painting curves can be called, displayed and saved.

___________________
*Work supported by National Natural Science Foundation of China (Project Nos. 11205185, 11175020 and 11175193)

[#]huangmy@ihep.ac.cn

## INJECTION BEAM CONTROL

For CSNS/RCS, the accuracy of all the 32 BPMs may not detect the single bunch beam. However, in order to calculate the matching position and direction of the injection beam, there are two special BPMs (INBPM01 and INBPM02) whose accuracy is higher than the ordinary BPMs of RCS and can be used to measure the positions of the single bunch beam. Figure 3 shows the control interface of the single bunch beam calculation. From this figure, if the position of single bunch beam is nearly unchanged for several turns, it can be known that the position and direction of injection beam are matching. Oppositely, the position and direction of injection beam are not suitable. With the transmission matrix and the measured values of INBPM01 and INBPM02, the correct position and direction of injection beam can be calculated.

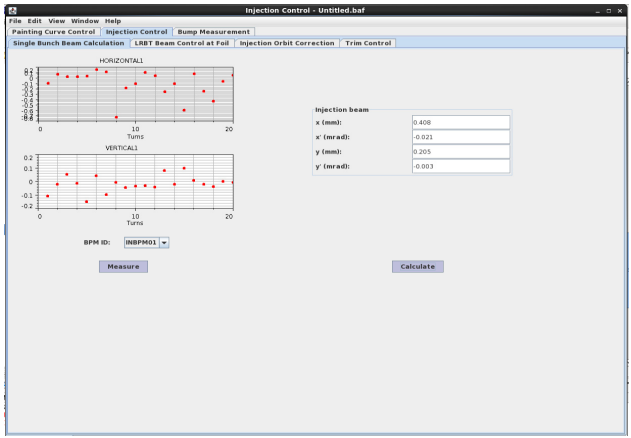

Figure 3: Single-bunch-beam calculation.

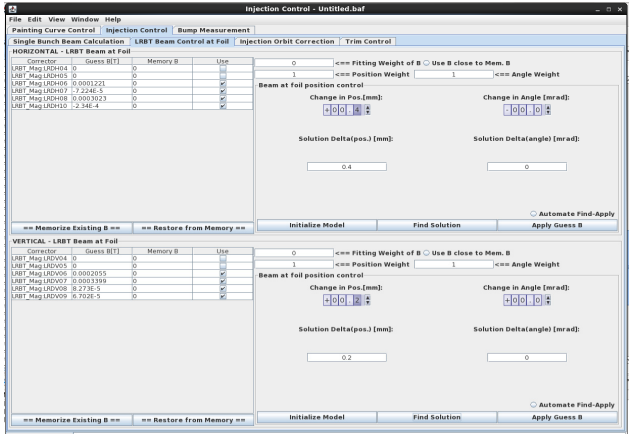

Figure 4: LRBT beam control at the foil.

While the matching position and direction of injection beam are obtained, with the correctors of LRBT, the injection beam can be corrected to fit for the RCS orbit. Figure 4 shows the control interface that the injection beam adjusted by the LRBT correctors at the foil.

## INJECTION ORBIT CORRECTION

Theoretically, the four BC magnets can produce an ideal bump and make no difference to the ring orbit. However, if the injection orbit with BC bump gives away, it will need to be corrected by the trim powers of BC magnets. Figure 5 shows the control interface of the injection orbit correction. It can be known that, if the injection orbit gives away, with the transmission matrix, the optimized solutions of the three trim powers of BC magnets can be calculated. Then, the three trim powers can be set with the suitable values and the injection orbit can be corrected.

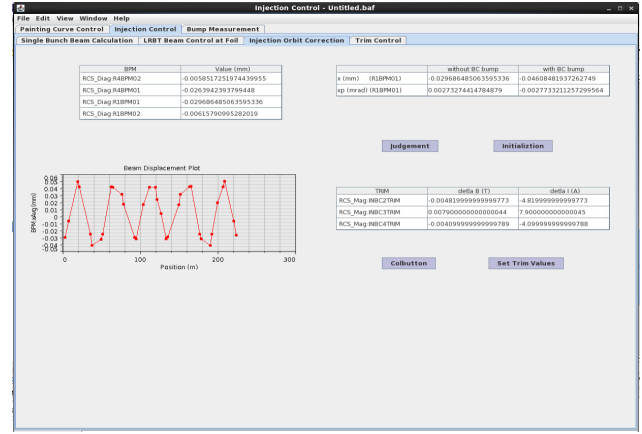

Figure 5: Injection orbit correction by the calculation.

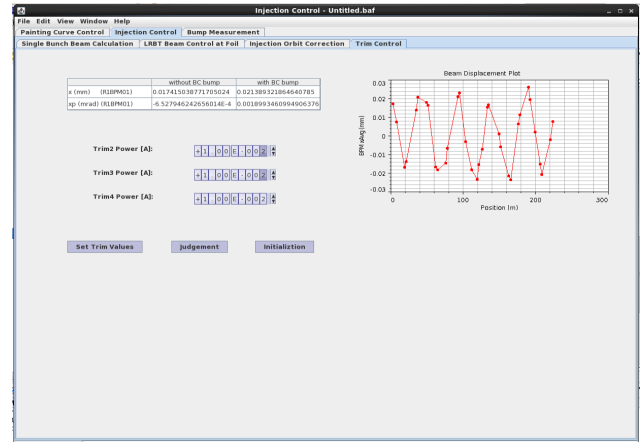

Figure 6: Injection trim power control.

However, if the suitable solutions of the trim powers cannot be given by the above program, it will be need to sharp tuning the three trim powers respectively. Figure 6 shows the control interface that the injection orbit controlled by the three trim powers.

## CONCLUSIONS

In this paper, the injection beam commissioning software for CSNS/RCS was introduced, and it contains three parts currently: painting curve control, injection beam control and injection orbit correction. The injection beam control contains two subsections: single bunch beam calculation and LRBT beam control at the foil. The injection orbit correction also contains two subsections: injection orbit correction by the calculation and injection trim power control.

## ACKNOWLEDGMENTS

The authors want to thank CSNS colleagues for the discussion and consultations.